\documentclass[prl,aps,twocolumn,showpacs]{revtex4}
\usepackage{graphicx}
\usepackage{amsmath}
\usepackage{bm}
\usepackage{amstext}
\usepackage{amsxtra}

\usepackage{times}

\begin{document}

\newcommand{\To}{T_c^0}
\newcommand{\kB}{k_{\rm B}}
\newcommand{\dT}{\Delta T_c}
\newcommand{\lo}{\lambda_0}
\newcommand{\cs}{$\clubsuit \; $}
\newcommand{\thold}{t_{\rm hold}}
\newcommand{\Nmf}{N_c^{\rm MF}}
\newcommand{\LM}{L_3^{\rm M}}
\newcommand{\Tmf}{T_c^{\rm MF}}
\newcommand{\bra}[1]{\langle #1|}
\newcommand{\ket}[1]{|#1\rangle}
\newcommand{\downstate}{\left\vert\downarrow\right\rangle}
\newcommand{\upstate}{\left\vert\uparrow\right\rangle}
\newcommand{\Ndown}{N_{\downarrow}}
\newcommand{\Nup}{N_{\uparrow}}

\title{
Stability of a unitary Bose gas
}

\author{Richard J. Fletcher, Alexander L. Gaunt, Nir Navon, Robert P. Smith*\email{rps24@cam.ac.uk}, and Zoran Hadzibabic}
\affiliation{Cavendish Laboratory, University of Cambridge, J.~J.~Thomson Avenue, Cambridge CB3~0HE, United Kingdom}

\begin{abstract}
We study the stability of a thermal $^{39}$K Bose gas across a broad Feshbach resonance, focusing on the unitary regime, where the scattering length $a$ exceeds the thermal wavelength $\lambda$.
We measure the general scaling laws relating the particle-loss and heating rates to the temperature, scattering length, and atom number. Both at unitarity and for positive $a \ll \lambda$ we find agreement with three-body theory. However, for $a<0$ and away from unitarity, we observe significant four-body decay. 
At unitarity, the three-body loss coefficient, $L_3 \propto \lambda^4$, is three times lower than the universal theoretical upper bound. This reduction is a consequence of species-specific Efimov physics and makes $^{39}$K particularly promising for studies of many-body physics in a unitary Bose gas.
\end{abstract}

\date{\today}


\maketitle

The control of interactions provided by Feshbach resonances makes ultracold atomic gases appealing for studies of both few- and many-body physics. On resonance, the s-wave scattering length $a$, which characterises two-body interactions, diverges. At and near the resonance a gas is in the unitary regime, where the interactions do not explicitly depend on the diverging $a$. Instead, $a$ is replaced by another natural lengthscale. In a degenerate gas this lengthscale is set by the inter-particle spacing; in a thermal gas it is set by the thermal wavelength $\lambda = h/\sqrt{2\pi m \kB T}$, where $m$ is the particle mass and $T$ is the temperature.

Over the past decade, there have been many studies of the unitary Fermi gas~\cite{Zwerger:2011}. More recently, there has been an increasing interest in both universal and species-specific properties of a unitary Bose gas~\cite{Cowell:2002,Song:2009,Lee:2010,Navon:2011,Borzov:2012,Li:2012,Wild:2012,Werner:2012,Rem:2013,Pricoupenko:2013,Castin:2013,vanHeugten:2013a,vanHeugten:2013b}. 
It is however an open question to what extent this state can be studied in (quasi-)equilibrium, since at unitarity three-body recombination leads to significant particle loss and heating~\cite{Pauli}. The severity of this instability is not universal~\cite{Rem:2013}, as it depends on the species-specific few-body Efimov physics~\cite{Kraemer:2006,Braaten:2007,Hammer:2007,Braaten:2008,Ferlaino:2009,vonStecher:2009,Gross:2009,Zaccanti:2009,Pollack:2009,Ferlaino:2011,Wild:2012,Roy:2013}.
Characterising and understanding the stability of a unitary Bose gas is thus important both from the perspective of Efimov physics and for identifying suitable atomic species for many-body experiments.

The per-particle loss rate due to three-body recombination is given by
\begin{equation}
\gamma_3 \equiv -\dot{N}/N = L_3\langle n^2\rangle  ,
\label{eq:L3}
\end{equation}
where $N$ is the atom number, $L_3$ is the three-body loss coefficient, $n$ is the density, and $\langle ... \rangle$ denotes an average over the density distribution in a trapped gas.
Away from unitarity, $L_3 \sim \hbar a^4/m$~\cite{Fedichev:1996d, Weber:2003}, with a dimensionless prefactor exhibiting additional variation with $a$ due to Efimov physics~\cite{Braaten:2007,Ferlaino:2011}. 
At unitarity
$L_3$ should saturate at $\sim \hbar \lambda^4/m \propto 1/T^2$. Experimental evidence for such saturation was observed in~\cite{Kraemer:2006,Wild:2012,Rem:2013}. More quantitatively, at unitarity we expect
\begin{equation}
\label{eq:C}
L_3\approx \zeta  \frac{9\sqrt{3}\hbar}{m}\lambda^4  = \zeta \frac{36 \sqrt{3} \pi^2 \hbar^5}{m^3 (\kB T)^2}   ,
\end{equation}
where $\zeta \leq 1$ is a species-dependent, non-universal dimensionless constant~\cite{Rem:2013} (see also~\cite{Greene:2004,DIncao:2004,Mehta:2009}).

Similar scaling arguments apply to the two-body elastic scattering rate, $\gamma_2$, which drives continuous re-equilibration of the gas during loss and heating. Away from unitarity $\gamma_2 \propto  \langle n \rangle \hbar a^2/(m\lambda)$; hence, at unitarity $\gamma_2 \propto \langle n \rangle \hbar \lambda/m$. 
The possibility to experimentally explore many-body physics of a quasi-equilibrium unitary Bose gas depends on the ratio $\gamma_3/\gamma_2$. Remarkably, at a given phase-space density, $n\lambda^3$, this ratio depends only on the species-specific $\zeta$.

Recently, $\zeta\approx 0.9$ was measured for $^7$Li~\cite{Rem:2013}. The gas was held in a relatively shallow trap, so that continuous evaporation converted heating into an additional particle loss, and the extraction of $\zeta$ relied on theoretically modelling this conversion and assuming the $1/T^2$ scaling of Eq.~(\ref{eq:C}).

In this Letter, we study the stability of the $^{39}$K Bose gas in the $|F, m_F\rangle = |1,1\rangle$ hyperfine ground state, across a broad Feshbach resonance centred at $402.5$ G~\cite{Zaccanti:2009}. 
We perform experiments in a deep trap and verify the predicted recombination-heating rate both at unitarity and for positive $a \ll \lambda$~\cite{Rem:2013,Weber:2003}. At unitarity we measure $L_3 \propto T^{-1.7\pm0.3}$  and $\zeta \approx 0.3$, a value that makes $^{39}$K particularly promising for studies of an equilibrium unitary gas. 
Additional measurements at $a<0$, away from unitarity, reveal the importance of four-body processes~\cite{Hammer:2007,vonStecher:2009}, consistent with previous studies in $^{133}$Cs~\cite{Ferlaino:2009}, $^{39}$K~\cite{Zaccanti:2009}, and $^7$Li~\cite{Pollack:2009}. 

Our experimental setup is described in Ref.~\cite{Campbell:2010}. We start by preparing a weakly interacting ($\lambda/a \approx 35$) thermal gas  in a harmonic optical trap. The trap has a depth  of $U \approx \kB \times 30\;\mu$K and is nearly isotropic, with the geometric mean of the trapping frequencies $\omega=2\pi\times 185\;$Hz. We then tune $a$ close to a Feshbach resonance, by ramping an external magnetic field over $10$~ms. At this point we have $N \approx 10^5$ atoms at $T \approx 1\;\mu$K, corresponding to $\lambda \approx 5 \times 10^3\;a_0$, where $a_0$ is the Bohr radius. 
At the trap centre $n \approx 3 \times 10^{12}\,$cm$^{-3}$ and $n\lambda^3<0.1$, so even at unitarity and assuming $\zeta=1$, we still always have $\gamma_2\gg\gamma_3$. 
We let the cloud evolve for a variable hold time, $t$, of up to 4~s, and then simultaneously switch off the trap and the Feshbach field (within $\sim 100\;\mu$s~\cite{molecules}). 
Finally, we image the cloud after 5 ms of time-of-flight expansion.

\begin{figure}[tbp] 
 \includegraphics[width=1\columnwidth]{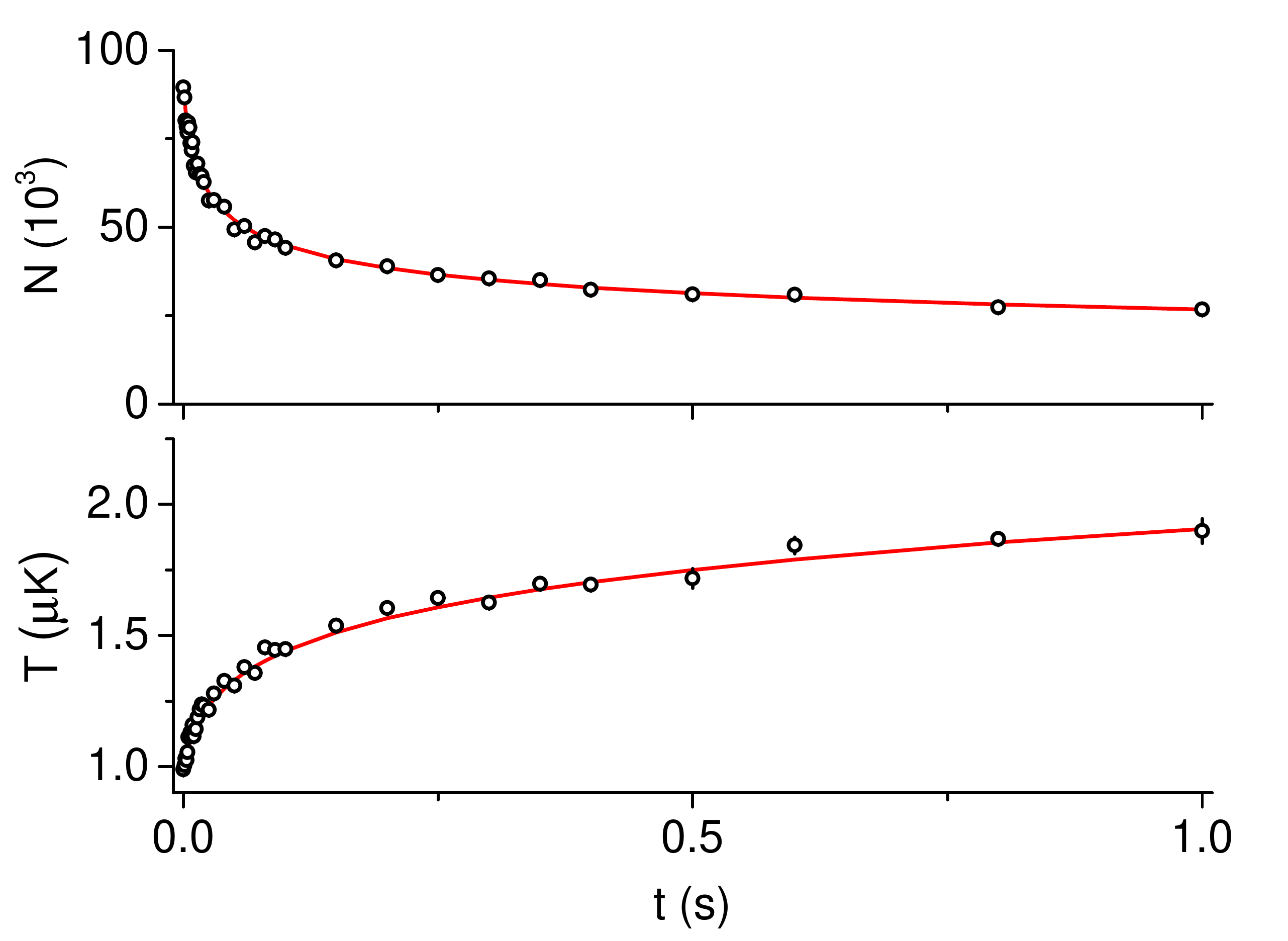}
 \caption{(color online) Particle loss and heating in a resonantly-interacting Bose gas ($\lambda/a=0$). Each point is an average of 5 measurements and error bars show standard statistical errors. Solid red lines are fits based on Eqs.~(\ref{eq:model}) and (\ref{eq:beta}). }
 \label{fig:decay}
\end{figure}

Fig.~\ref{fig:decay} shows the particle loss and heating in a resonantly-interacting gas ($\lambda/a=0$). 
Restricting our measurements to $T < 2\;\mu$K ensures that evaporative losses and cooling are negligible. 
We have taken 19 similar data series, each at a fixed $a$, spanning the range $-12 < \lambda/a < 12$.

We first study the relationship between $T$ and $N$ during the evolution of the cloud.
One expects three sources of heating related to three-body recombination~\cite{Weber:2003,Rem:2013}: (i) For any $a$, losses preferentially occur near the centre of the cloud, where the atoms have lower potential energy. 
(ii) For $a>0$, recombination results in a shallow dimer with binding energy $\varepsilon = \hbar^2/(ma^2)$, and the third atom carries away $(2/3) \varepsilon$ as kinetic energy. In all our experiments $\varepsilon <  U$, so this atom remains trapped and increases the energy of the cloud.
(iii) At unitarity, three-body recombination preferentially involves atoms that also have lower kinetic energy.

\begin{figure}[tbp]
   \centering
   \includegraphics[width=\columnwidth]{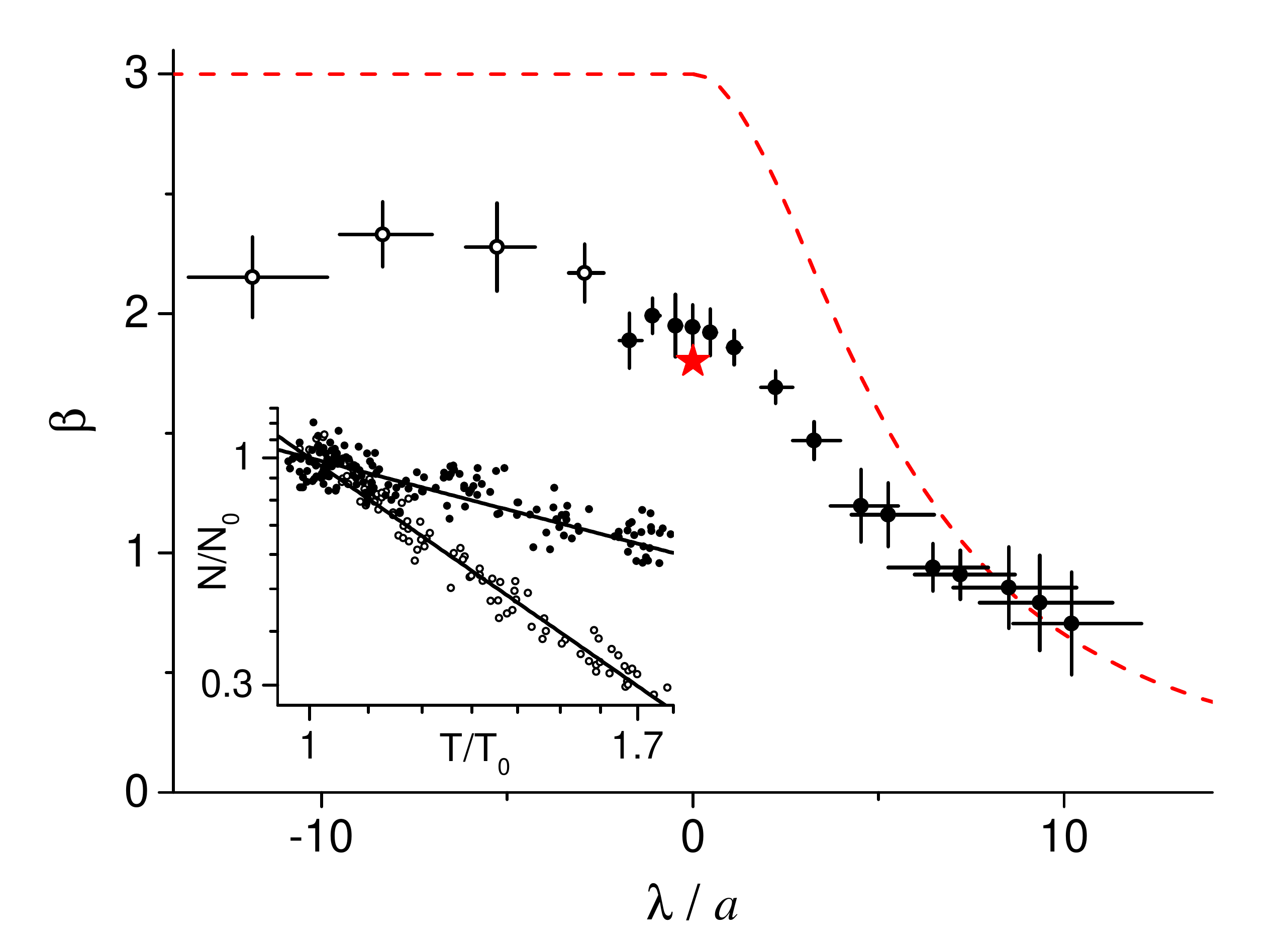} 
   \caption{(color online) Heating exponent $\beta$, as defined in Eq.~(\ref{eq:beta}). The red dashed line is a result of non-unitary three-body theory, while the red star indicates the predicted value of 1.8 at unitarity. Open symbols indicate the region where four-body decay is significant (see text and Fig~\ref{fig:4body}). Note that $\lambda \approx 5\times 10^3 a_0$ and horizontal error bars reflect its variation during a measurement sequence at a fixed $a$.  Vertical error bars show fitting uncertainties.
Inset: log-log plots of $N$ vs. $T$ (scaled to their values at $t=0$) for the data series at $\lambda/a \approx -5.3$ (open) and $8.5$ (solid).}
   \label{fig:beta}
\end{figure}

To a good approximation, in our experiments we can capture all these effects by a simple scaling law:
\begin{equation}
\label{eq:beta}
NT^{\beta} = {\rm const.}  ,
\end{equation}
with the exponent $\beta$ varying across the resonance.
Ignoring unitarity effects, $\beta =3$ for $a \leq 0$, and  $\beta = 3/[1+ \lambda^2/(9\pi a^2)]$ for $a>0$ (see also~\cite{Weber:2003}). In the latter case $\beta$ changes as the cloud heats, but in our measurements this variation is small enough that a constant $\beta = - d[\ln(N)]/d[\ln(T)]$ describes the data well (see inset of Fig.~\ref{fig:beta}). At unitarity, a universal value of $\beta=1.8$ was predicted in Ref.~\cite{Rem:2013}.


In Fig.~\ref{fig:beta} we show our measured values of $\beta$. 
For  $\lambda/a \gg 1$ 
we find agreement with the non-unitary prediction shown by the red dashed line. However, approaching unitarity we see gradual deviation from this theory. On resonance, we measure $\beta=1.94\pm0.09$, close to the unitary prediction of $\beta=1.8$ (indicated by the red star), and far from the non-unitary $\beta=3$.

\begin{figure}[tbp]
   \centering
   \includegraphics[width=\columnwidth]{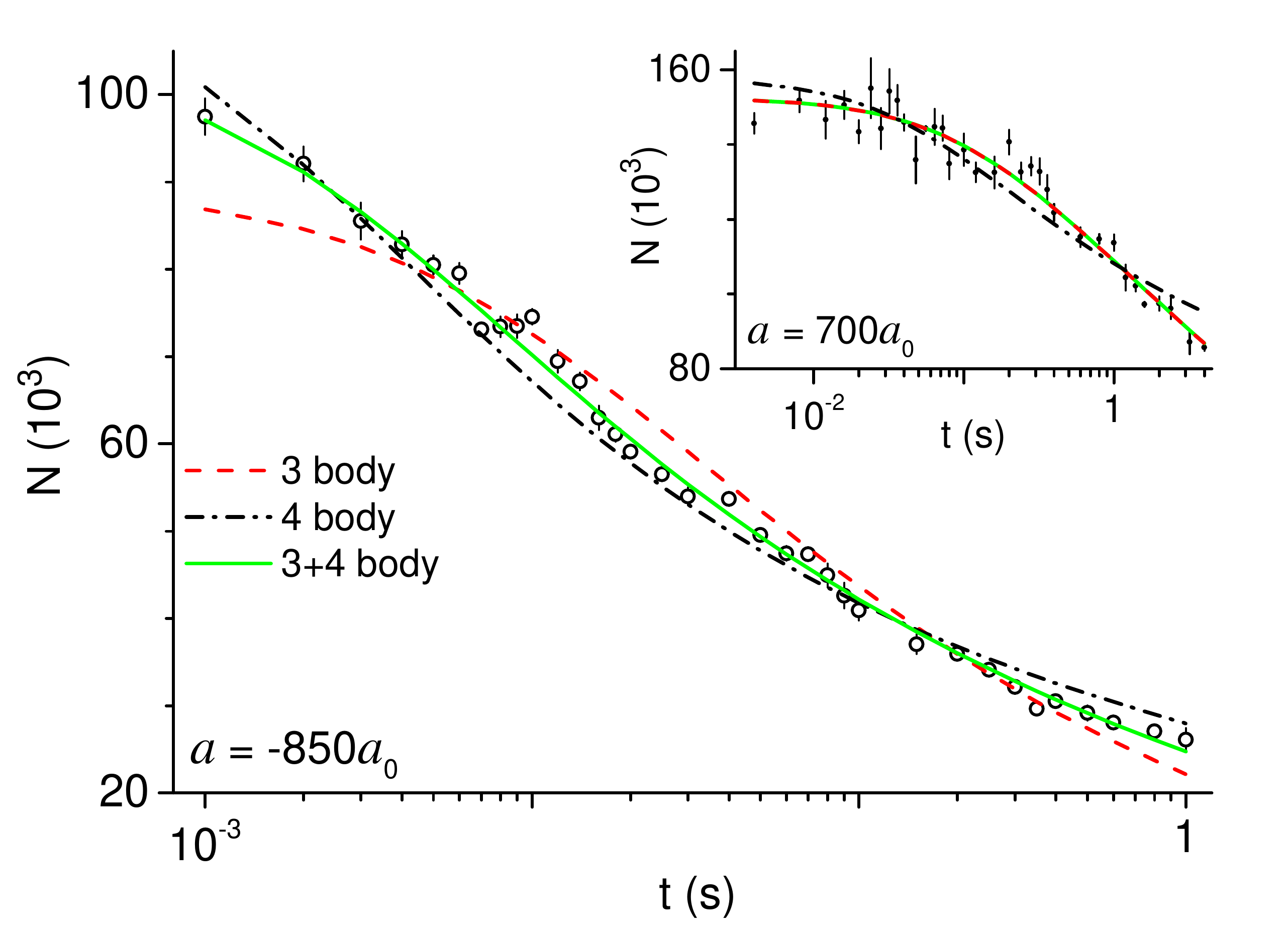} 
   \caption{(color online) Three- vs. four-body decay for $a<0$ (away from unitarity). 
$N$-decay at $a=-850 \; a_0$ is fitted to a model including both three- and four-body losses (green solid line), as well as to pure three- and four-body models (red dashed and black dot-dashed line, respectively). 
Inset: for comparison, at $a= 700 \; a_0$, the solid green and the dashed red lines are indistinguishable, showing that four-body decay does not play a detectable role.}
   \label{fig:4body}
\end{figure}

Moving away from unitarity into the $a<0$ region (open symbols in Fig.~\ref{fig:beta}, corresponding to $-2000 < a/a_0 < -400$), $\beta$ rises further, but does not reach the expected non-unitary limit. 
By analysing the dynamics of the particle loss, $N(t)$, we find that in this region four-body decay
is also significant (see Fig.~\ref{fig:4body}); in this case our prediction for $\beta$ is not applicable.
Previously, indirect evidence for four-body decay in this region was seen in Ref.~\cite{Zaccanti:2009}, but not in Ref.~\cite{Roy:2013}, where the initial cloud density was significantly lower.

We fit the $N(t)$ data by numerically evolving a loss equation featuring both three- and four-body decay~\cite{Ferlaino:2009},
\begin{equation}
\label{eq:numerical}
\dot{N}=-L_3\langle n^2\rangle N-L_4 \langle n^3 \rangle N ,
\end{equation}
where $L_3$ and $L_4$ are fitting parameters and we use the measured $T(t)$ to evaluate the thermal density averages. 
To obtain purely three- (four-) body fits we fix $L_4$ ($L_3$) to zero. 

In Fig.~\ref{fig:4body} we show $N(t)$ for $a = -850\;a_0$. The model including both $L_3$ and $L_4$ provides an excellent fit to the data, the pure four-body fit is comparable, while the pure three-body fit is quite poor. 
We observe  four-body effects for all our data with $-2000 < a/a_0 < -400$. However, we find that they are relevant only at densities $\gtrsim 10^{12}\,$cm$^{-3}$, which reconciles the observations of Refs.~\cite{Zaccanti:2009} and~\cite{Roy:2013}. A more detailed study of this region, including any four-body resonances~\cite{Ferlaino:2009}, is outside the scope of this paper.

For $a>0$ the same analysis does not reveal any four-body decay (see inset of Fig.~\ref{fig:4body}). 
In this case the pure three-body fit and the fit including both $L_3$ and $L_4$ are indistinguishable, and give the same $L_3$ (within the $10\%$ fitting errors). 
This strongly excludes $L_4$ as a relevant fit parameter. 
Using a similar procedure, we have also checked that for both positive and negative $a$ we do not detect any five-body decay.

We henceforth focus on the three-body decay dynamics at unitarity, using the $a>0$ non-unitary regime for comparison.
Invoking  Eq.~(\ref{eq:beta}), in both regimes the particle loss should be described by:
\begin{equation}
\dot{N}=-AN^\nu  ,
\label{eq:model}
\end{equation}
where $A$ and $\nu$ are constants. Here, $\nu$ absorbs all the $N$ and $T$ dependence of $L_3$ and $\langle n^2 \rangle$. Integration gives a fitting function $ N(t) = \left[ A(\nu - 1) t + N(0)^{1-\nu} \right]^{1/(1-\nu)}$.
For $a \ll \lambda$ we expect $\nu=3+3/\beta$, whereas at unitarity $L_3 \propto 1/T^2$ implies $\nu=3+5/\beta$. To test this hypothesis in an unbiased way, we analyse our data using $\nu$ as a free parameter. 

Note that here we invoke Eq.~(\ref{eq:beta}) merely to anticipate the validity of Eq.~(\ref{eq:model}) and the $\nu$ values; experimentally, our analysis of $N(t)$ and $\nu$ is decoupled from the measurements of $T(t)$ and $\beta$.
The validity of our approach is seen in Fig.~\ref{fig:decay}, where the fit of $N(t)$ is based on Eq.~(\ref{eq:model}). The fit of $T(t)$ is then obtained by inserting the fitted $N(t)$ and $\beta$  into Eq.~(\ref{eq:beta}). 

\begin{figure}[tbp]
   \includegraphics[width=1\columnwidth]{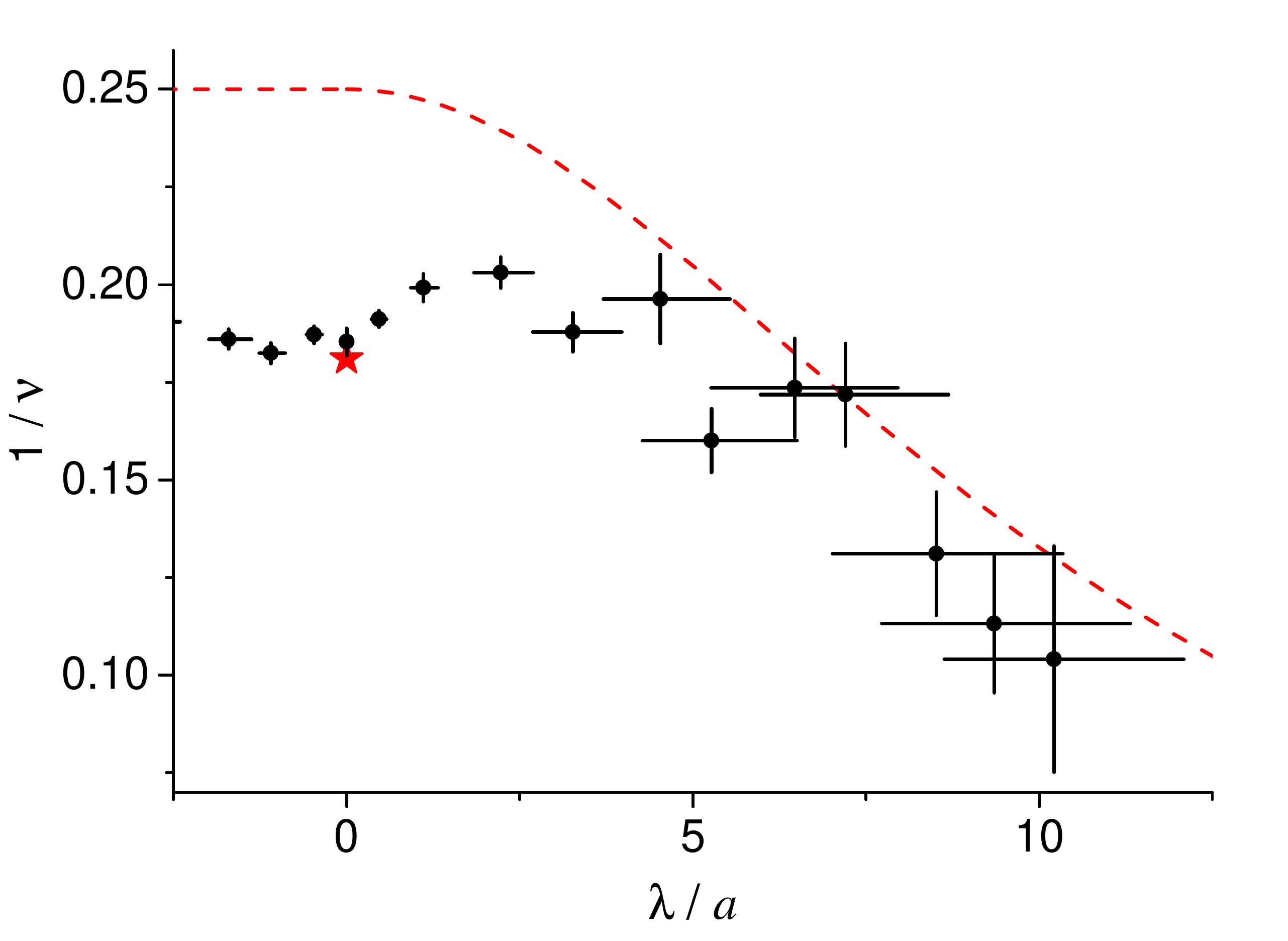} 
   \caption{(color online) Particle-loss exponent $\nu$, as defined in Eq.~(\ref{eq:model}). The red dashed line shows the non-unitary theory, $\nu = 3 + 3/\beta$, assuming non-unitary $\beta$ values. The red star shows the unitary prediction, $\nu = 3 + 5/\beta$, corresponding to $L_3\propto 1/T^2$ and the measured $\beta$. Error bars are analogous to those in Fig.~\ref{fig:beta}.}
   \label{fig:gamma}
\end{figure}

Our fitted values of $\nu$ are summarised in Fig.~\ref{fig:gamma}. We see a crossover from non-unitary to unitary behaviour as the resonance is approached, confirming the appearance of a temperature-dependent $L_3$. 
Now combining our measurements of $\beta$ and $\nu$, at unitarity we get $L_3\propto T^{-1.7\pm0.3}$, in agreement with the expected $1/T^2$ scaling. 

Next, using the fitted $A$ and $\nu$, for each data series at a particular $a$, and for any evolution time $t$, we extract:
\begin{equation}
L_3(t) = 3\sqrt{3}  \left(  \frac{2\pi k_BT(t)}{m \omega^2} \right)^{3} N(t)^{\nu-3} A \,  .
\label{eq:fullL3}
\end{equation}
Combining all our data series, we reconstruct $L_3(a,T)$. 

\begin{figure}[tbp]
   \includegraphics[width=1\columnwidth]{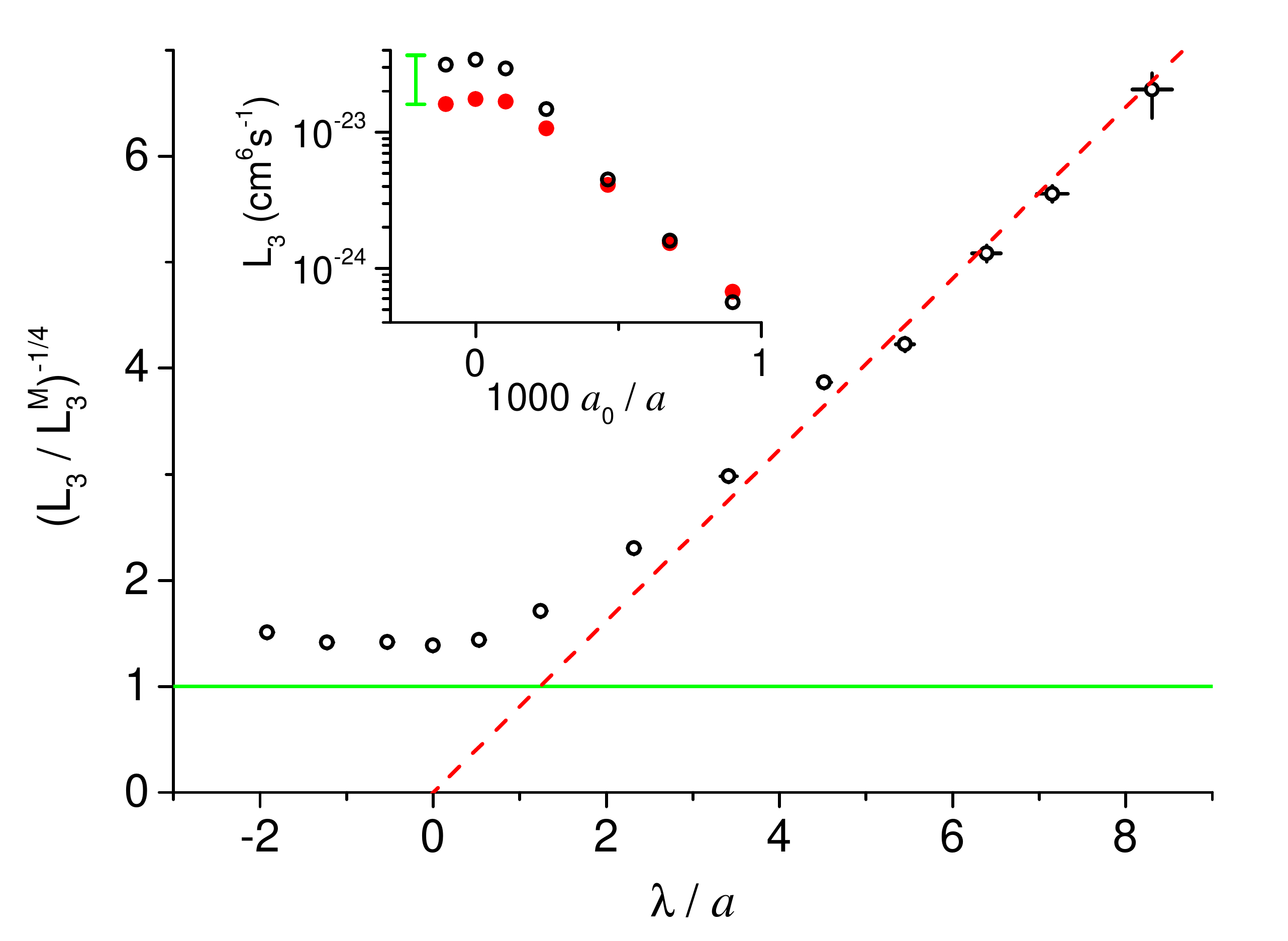} 
   \caption{(color online) Three-body loss coefficient. Main panel: $(L_3/\LM)^{-1/4}$ (see text) at $T=1.1\;\mu$K. Horizontal green line marks the theoretical upper bound on $L_3$, while the red dashed line is a guide to the eye showing the $ L_3 \propto a^4$ non-unitary scaling. At unitarity, $L_3/\LM \approx 0.27$. Inset: $L_3$ at $1.1\;\mu$K (open symbols) and $1.7\;\mu$K (solid symbols). The expected ratio between the two unitary plateaux is indicated by the green vertical bar.}
   \label{fig:L3}
\end{figure}

In Fig.~\ref{fig:L3} (main panel) we show $L_3$ at a fixed $T=1.1 \;\mu$K, scaled to the theoretical upper bound $\LM(T)$, obtained by setting $\zeta = 1$ in Eq.~(\ref{eq:C}). Plotting 
$(L_3/\LM)^{-1/4}$ versus $\lambda/a$ clearly reveals two key effects. First, for $\lambda/a \gtrsim 3$, we see the non-unitary scaling $L_3 \propto a^4$~\cite{no_wiggles}. Second, close to the resonance, $L_3$ saturates at $\approx 0.27 \, \LM$.

In the inset of Fig.~\ref{fig:L3} we focus on the region close to the resonance and compare $L_3$ for two different temperatures, $T=1.1\;\mu$K and $1.7\;\mu$K. Away from the resonance, $L_3$ does not show any $T$-dependence. At unitarity, the ratio of the two saturated $L_3$ values is close to the expected $1/T^2$ scaling.

Finally, to refine our estimate of $\zeta$, we fix $\nu=3+5/\beta$ (i.e., $L_3 \propto 1/T^2$) and reanalyse
the three data series taken closest to the resonance, for which $|\lambda/a|<0.6$ at all times. This gives us a combined estimate of $\zeta=0.29 \pm 0.03$, while the systematic uncertainty in $\zeta$ due to our absolute atom-number calibration~\cite{Smith:2011, Smith:2011b} is about $30\%$.
Writing $L_3 = \lambda_3/ T^2$, this corresponds to $\lambda_3 \approx 4.5 \times 10^{-23}\;(\mu$K$)^2$cm$^6$s$^{-1}$.
In the context of Efimov physics, $\zeta = 1- e^{-4\eta}$, where $\eta$ is the Efimov width parameter~\cite{Braaten:2003}. We deduce $\eta = 0.09 \pm 0.04$ (see also~\cite{Zaccanti:2009}).

In conclusion, we have fully characterised the stability of a $^{39}$K gas at and near unitarity. We have experimentally verified the theoretically predicted general scaling laws characterising particle loss and heating in the unitary regime, confirmed the relevance of four-body decay on the negative side of the Feshbach resonance, and measured the species-specific unitarity-limited three-body loss coefficient, $L_3 \propto 1/T^2$. The unitary value of $L_3$, three times lower than the universal theoretical upper bound, makes $^{39}$K a promising candidate for experimental studies of many-body physics in a unitary Bose gas.

We thank F. Chevy, J. Dalibard, A. Grier, P. Massignan, B. Rem, and F. Werner for useful discussions and comments on the manuscript.
This work was supported by EPSRC (Grant No. EP/K003615/1), the Royal Society, AFOSR, ARO and DARPA OLE.


\begin{thebibliography}{36}

\item[*]{e-mail: rps24@cam.ac.uk}\vspace{2mm}

\expandafter\ifx\csname natexlab\endcsname\relax\def\natexlab#1{#1}\fi
\expandafter\ifx\csname bibnamefont\endcsname\relax
  \def\bibnamefont#1{#1}\fi
\expandafter\ifx\csname bibfnamefont\endcsname\relax
  \def\bibfnamefont#1{#1}\fi
\expandafter\ifx\csname citenamefont\endcsname\relax
  \def\citenamefont#1{#1}\fi
\expandafter\ifx\csname url\endcsname\relax
  \def\url#1{\texttt{#1}}\fi
\expandafter\ifx\csname urlprefix\endcsname\relax\def\urlprefix{URL }\fi
\providecommand{\bibinfo}[2]{#2}
\providecommand{\eprint}[2][]{\url{#2}}


\bibitem[{\citenamefont{Zwerger}(2011)}]{Zwerger:2011}
\bibinfo{editor}{\bibfnamefont{W.}~\bibnamefont{Zwerger}}, ed.,
  \emph{\bibinfo{title}{BCS-BEC Crossover and the Unitary Fermi Gas,}}, vol.
  \bibinfo{volume}{836} of \emph{\bibinfo{series}{Lecture Notes in Physics}}
  (\bibinfo{publisher}{Springer}, \bibinfo{address}{Berlin},
  \bibinfo{year}{2011}).

\bibitem[{\citenamefont{Cowell et~al.}(2002)\citenamefont{Cowell, Heiselberg,
  Mazets, Morales, Pandharipande, and Pethick}}]{Cowell:2002}
\bibinfo{author}{\bibfnamefont{S.}~\bibnamefont{Cowell}},
  \bibinfo{author}{\bibfnamefont{H.}~\bibnamefont{Heiselberg}},
  \bibinfo{author}{\bibfnamefont{I.~E.} \bibnamefont{Mazets}},
  \bibinfo{author}{\bibfnamefont{J.}~\bibnamefont{Morales}},
  \bibinfo{author}{\bibfnamefont{V.~R.} \bibnamefont{Pandharipande}},
  \bibnamefont{and} \bibinfo{author}{\bibfnamefont{C.~J.}
  \bibnamefont{Pethick}}, \bibinfo{journal}{Phys. Rev. Lett.}
  \textbf{\bibinfo{volume}{88}}, \bibinfo{pages}{210403}
  (\bibinfo{year}{2002}).

\bibitem[{\citenamefont{Song and Zhou}(2009)}]{Song:2009}
\bibinfo{author}{\bibfnamefont{J.~L.} \bibnamefont{Song}} \bibnamefont{and}
  \bibinfo{author}{\bibfnamefont{F.}~\bibnamefont{Zhou}},
  \bibinfo{journal}{Phys. Rev. Lett.} \textbf{\bibinfo{volume}{103}},
  \bibinfo{pages}{025302} (\bibinfo{year}{2009}).

\bibitem[{\citenamefont{Lee and Lee}(2010)}]{Lee:2010}
\bibinfo{author}{\bibfnamefont{Y.-L.} \bibnamefont{Lee}} \bibnamefont{and}
  \bibinfo{author}{\bibfnamefont{Y.-W.} \bibnamefont{Lee}},
  \bibinfo{journal}{Phys. Rev. A} \textbf{\bibinfo{volume}{81}},
  \bibinfo{pages}{063613} (\bibinfo{year}{2010}).

\bibitem[{\citenamefont{Navon et~al.}(2011)\citenamefont{Navon, Piatecki,
  G\"unter, Rem, Nguyen, Chevy, Krauth, and Salomon}}]{Navon:2011}
\bibinfo{author}{\bibfnamefont{N.}~\bibnamefont{Navon}},
  \bibinfo{author}{\bibfnamefont{S.}~\bibnamefont{Piatecki}},
  \bibinfo{author}{\bibfnamefont{K.}~\bibnamefont{G\"unter}},
  \bibinfo{author}{\bibfnamefont{B.}~\bibnamefont{Rem}},
  \bibinfo{author}{\bibfnamefont{T.~C.} \bibnamefont{Nguyen}},
  \bibinfo{author}{\bibfnamefont{F.}~\bibnamefont{Chevy}},
  \bibinfo{author}{\bibfnamefont{W.}~\bibnamefont{Krauth}}, \bibnamefont{and}
  \bibinfo{author}{\bibfnamefont{C.}~\bibnamefont{Salomon}},
  \bibinfo{journal}{Phys. Rev. Lett.} \textbf{\bibinfo{volume}{107}},
  \bibinfo{pages}{135301} (\bibinfo{year}{2011}).

\bibitem[{\citenamefont{Borzov et~al.}(2012)\citenamefont{Borzov, Mashayekhi,
  Zhang, Song, and Zhou}}]{Borzov:2012}
\bibinfo{author}{\bibfnamefont{D.}~\bibnamefont{Borzov}},
  \bibinfo{author}{\bibfnamefont{M.~S.} \bibnamefont{Mashayekhi}},
  \bibinfo{author}{\bibfnamefont{S.}~\bibnamefont{Zhang}},
  \bibinfo{author}{\bibfnamefont{J.-L.} \bibnamefont{Song}}, \bibnamefont{and}
  \bibinfo{author}{\bibfnamefont{F.}~\bibnamefont{Zhou}},
  \bibinfo{journal}{Phys. Rev. A} \textbf{\bibinfo{volume}{85}},
  \bibinfo{pages}{023620} (\bibinfo{year}{2012}).

\bibitem[{\citenamefont{Li and Ho}(2012)}]{Li:2012}
\bibinfo{author}{\bibfnamefont{W.}~\bibnamefont{Li}} \bibnamefont{and}
  \bibinfo{author}{\bibfnamefont{T.-L.} \bibnamefont{Ho}},
  \bibinfo{journal}{Phys. Rev. Lett.} \textbf{\bibinfo{volume}{108}},
  \bibinfo{pages}{195301} (\bibinfo{year}{2012}).

\bibitem[{\citenamefont{Wild et~al.}(2012)\citenamefont{Wild, Makotyn, Pino,
  Cornell, and Jin}}]{Wild:2012}
\bibinfo{author}{\bibfnamefont{R.~J.} \bibnamefont{Wild}},
  \bibinfo{author}{\bibfnamefont{P.}~\bibnamefont{Makotyn}},
  \bibinfo{author}{\bibfnamefont{J.~M.} \bibnamefont{Pino}},
  \bibinfo{author}{\bibfnamefont{E.~A.} \bibnamefont{Cornell}},
  \bibnamefont{and} \bibinfo{author}{\bibfnamefont{D.~S.} \bibnamefont{Jin}},
  \bibinfo{journal}{Phys. Rev. Lett.} \textbf{\bibinfo{volume}{108}},
  \bibinfo{pages}{145305} (\bibinfo{year}{2012}).

\bibitem[{\citenamefont{Werner and Castin}(2012)}]{Werner:2012}
\bibinfo{author}{\bibfnamefont{F.}~\bibnamefont{Werner}} \bibnamefont{and}
  \bibinfo{author}{\bibfnamefont{Y.}~\bibnamefont{Castin}},
  \bibinfo{journal}{Phys. Rev. A} \textbf{\bibinfo{volume}{86}},
  \bibinfo{pages}{053633} (\bibinfo{year}{2012}).

\bibitem[{\citenamefont{Rem et~al.}(2013)\citenamefont{Rem, Grier,
  Ferrier-Barbut, Eismann, Langen, Navon, Khaykovich, Werner, Petrov, Chevy
  et~al.}}]{Rem:2013}
\bibinfo{author}{\bibfnamefont{B.~S.} \bibnamefont{Rem}},
  \bibinfo{author}{\bibfnamefont{A.~T.} \bibnamefont{Grier}},
  \bibinfo{author}{\bibfnamefont{I.}~\bibnamefont{Ferrier-Barbut}},
  \bibinfo{author}{\bibfnamefont{U.}~\bibnamefont{Eismann}},
  \bibinfo{author}{\bibfnamefont{T.}~\bibnamefont{Langen}},
  \bibinfo{author}{\bibfnamefont{N.}~\bibnamefont{Navon}},
  \bibinfo{author}{\bibfnamefont{L.}~\bibnamefont{Khaykovich}},
  \bibinfo{author}{\bibfnamefont{F.}~\bibnamefont{Werner}},
  \bibinfo{author}{\bibfnamefont{D.~S.} \bibnamefont{Petrov}},
  \bibinfo{author}{\bibfnamefont{F.}~\bibnamefont{Chevy}},
  \bibnamefont{and}  \bibinfo{author}{\bibfnamefont{C.}~\bibnamefont{Salomon}},
  \bibinfo{journal}{Phys. Rev. Lett.}
  \textbf{\bibinfo{volume}{110}}, \bibinfo{pages}{163202}
  (\bibinfo{year}{2013}).

\bibitem[{\citenamefont{{Pricoupenko}}(2013)}]{Pricoupenko:2013}
\bibinfo{author}{\bibfnamefont{L.}~\bibnamefont{{Pricoupenko}}},
  \bibinfo{journal}{Phys. Rev. Lett.} \textbf{\bibinfo{volume}{110}},
  \bibinfo{eid}{180402} (\bibinfo{year}{2013}).

\bibitem[{\citenamefont{{Castin} and {Werner}}(2013)}]{Castin:2013}
\bibinfo{author}{\bibfnamefont{Y.}~\bibnamefont{{Castin}}} \bibnamefont{and}
  \bibinfo{author}{\bibfnamefont{F.}~\bibnamefont{{Werner}}},
  \bibinfo{journal}{Canadian Journal of Physics} \textbf{\bibinfo{volume}{91}},
  \bibinfo{pages}{382} (\bibinfo{year}{2013}).

\bibitem[{\citenamefont{{van Heugten} and
  {Stoof}}(2013{\natexlab{a}})}]{vanHeugten:2013a}
\bibinfo{author}{\bibfnamefont{J.~J.~R.~M.} \bibnamefont{{van Heugten}}}
  \bibnamefont{and} \bibinfo{author}{\bibfnamefont{H.~T.~C.}
  \bibnamefont{{Stoof}}}, \bibinfo{journal}{arXiv:1302.1792}.


\bibitem[{\citenamefont{{van Heugten} and
  {Stoof}}(2013{\natexlab{b}})}]{vanHeugten:2013b}
\bibinfo{author}{\bibfnamefont{J.~J.~R.~M.} \bibnamefont{{van Heugten}}}
  \bibnamefont{and} \bibinfo{author}{\bibfnamefont{H.~T.~C.}
  \bibnamefont{{Stoof}}}, \bibinfo{journal}{arXiv:1306.1104}.
  
\bibitem{Pauli}
This is in contrast to a two-component Fermi gas, where three-body effects are suppressed by Pauli exclusion~\cite{Petrov:2004b}.

\bibitem[{\citenamefont{Petrov et~al.}(2004)\citenamefont{Petrov, Salomon, and
  Shlyapnikov}}]{Petrov:2004b}
\bibinfo{author}{\bibfnamefont{D.~S.} \bibnamefont{Petrov}},
  \bibinfo{author}{\bibfnamefont{C.}~\bibnamefont{Salomon}}, \bibnamefont{and}
  \bibinfo{author}{\bibfnamefont{G.~V.} \bibnamefont{Shlyapnikov}},
  \bibinfo{journal}{Phys. Rev. Lett.} \textbf{\bibinfo{volume}{93}},
  \bibinfo{pages}{090404} (\bibinfo{year}{2004}).

\bibitem[{\citenamefont{Kraemer et~al.}(2006)\citenamefont{Kraemer, Mark,
  Waldburger, Danzl, Chin, Engeser, Lange, Pilch, Jaakkola, Naegerl
  et~al.}}]{Kraemer:2006}
\bibinfo{author}{\bibfnamefont{T.}~\bibnamefont{Kraemer}},
  \bibinfo{author}{\bibfnamefont{M.}~\bibnamefont{Mark}},
  \bibinfo{author}{\bibfnamefont{P.}~\bibnamefont{Waldburger}},
  \bibinfo{author}{\bibfnamefont{J.~G.} \bibnamefont{Danzl}},
  \bibinfo{author}{\bibfnamefont{C.}~\bibnamefont{Chin}},
  \bibinfo{author}{\bibfnamefont{B.}~\bibnamefont{Engeser}},
  \bibinfo{author}{\bibfnamefont{A.~D.} \bibnamefont{Lange}},
  \bibinfo{author}{\bibfnamefont{K.}~\bibnamefont{Pilch}},
  \bibinfo{author}{\bibfnamefont{A.}~\bibnamefont{Jaakkola}},
  \bibinfo{author}{\bibfnamefont{H.~C.} \bibnamefont{N{\"a}gerl}},
  \bibnamefont{and} \bibinfo{author}{\bibfnamefont{R.}~\bibnamefont{Grimm}},
  \bibinfo{journal}{Nature}
  \textbf{\bibinfo{volume}{440}}, \bibinfo{pages}{315} (\bibinfo{year}{2006}).

\bibitem[{\citenamefont{Braaten and Hammer}(2007)}]{Braaten:2007}
\bibinfo{author}{\bibfnamefont{E.}~\bibnamefont{Braaten}} \bibnamefont{and}
  \bibinfo{author}{\bibfnamefont{H.-W.} \bibnamefont{Hammer}},
  \bibinfo{journal}{Annals of Physics} \textbf{\bibinfo{volume}{322}},
  \bibinfo{pages}{120 } (\bibinfo{year}{2007}). 

\bibitem[{\citenamefont{Hammer and Platter}(2007)}]{Hammer:2007}
\bibinfo{author}{\bibfnamefont{H.-W.} \bibnamefont{Hammer}} \bibnamefont{and}
  \bibinfo{author}{\bibfnamefont{L.}~\bibnamefont{Platter}},
  \bibinfo{journal}{The European Physical Journal A}
  \textbf{\bibinfo{volume}{32}}, \bibinfo{pages}{113} (\bibinfo{year}{2007}).

\bibitem[{\citenamefont{Braaten et~al.}(2008)\citenamefont{Braaten, Hammer,
  Kang, and Platter}}]{Braaten:2008}
\bibinfo{author}{\bibfnamefont{E.}~\bibnamefont{Braaten}},
  \bibinfo{author}{\bibfnamefont{H.-W.} \bibnamefont{Hammer}},
  \bibinfo{author}{\bibfnamefont{D.}~\bibnamefont{Kang}}, \bibnamefont{and}
  \bibinfo{author}{\bibfnamefont{L.}~\bibnamefont{Platter}},
  \bibinfo{journal}{Phys. Rev. A} \textbf{\bibinfo{volume}{78}},
  \bibinfo{pages}{043605} (\bibinfo{year}{2008}).

\bibitem[{\citenamefont{Ferlaino et~al.}(2009)\citenamefont{Ferlaino, Knoop,
  Berninger, Harm, D'Incao, N\"agerl, and Grimm}}]{Ferlaino:2009}
\bibinfo{author}{\bibfnamefont{F.}~\bibnamefont{Ferlaino}},
  \bibinfo{author}{\bibfnamefont{S.}~\bibnamefont{Knoop}},
  \bibinfo{author}{\bibfnamefont{M.}~\bibnamefont{Berninger}},
  \bibinfo{author}{\bibfnamefont{W.}~\bibnamefont{Harm}},
  \bibinfo{author}{\bibfnamefont{J.~P.} \bibnamefont{D'Incao}},
  \bibinfo{author}{\bibfnamefont{H.-C.} \bibnamefont{N\"agerl}},
  \bibnamefont{and} \bibinfo{author}{\bibfnamefont{R.}~\bibnamefont{Grimm}},
  \bibinfo{journal}{Phys. Rev. Lett.} \textbf{\bibinfo{volume}{102}},
  \bibinfo{pages}{140401} (\bibinfo{year}{2009}).

\bibitem[{\citenamefont{{von Stecher} et~al.}(2009)\citenamefont{{von Stecher},
  {D'Incao}, and {Greene}}}]{vonStecher:2009}
\bibinfo{author}{\bibfnamefont{J.}~\bibnamefont{{von Stecher}}},
  \bibinfo{author}{\bibfnamefont{J.~P.} \bibnamefont{{D'Incao}}},
  \bibnamefont{and} \bibinfo{author}{\bibfnamefont{C.~H.}
  \bibnamefont{{Greene}}}, \bibinfo{journal}{Nature Physics}
  \textbf{\bibinfo{volume}{5}}, \bibinfo{pages}{417} (\bibinfo{year}{2009}).

\bibitem[{\citenamefont{Gross et~al.}(2009)\citenamefont{Gross, Shotan,
  Kokkelmans, and Khaykovich}}]{Gross:2009}
\bibinfo{author}{\bibfnamefont{N.}~\bibnamefont{Gross}},
  \bibinfo{author}{\bibfnamefont{Z.}~\bibnamefont{Shotan}},
  \bibinfo{author}{\bibfnamefont{S.}~\bibnamefont{Kokkelmans}},
  \bibnamefont{and}
  \bibinfo{author}{\bibfnamefont{L.}~\bibnamefont{Khaykovich}},
  \bibinfo{journal}{Phys. Rev. Lett.} \textbf{\bibinfo{volume}{103}},
  \bibinfo{pages}{163202} (\bibinfo{year}{2009}).

\bibitem[{\citenamefont{Zaccanti et~al.}(2009)\citenamefont{Zaccanti, Deissler,
  D'Errico, Fattori, Jona-Lasinio, Mueller, Roati, Inguscio, and
  Modugno}}]{Zaccanti:2009}
\bibinfo{author}{\bibfnamefont{M.}~\bibnamefont{Zaccanti}},
  \bibinfo{author}{\bibfnamefont{B.}~\bibnamefont{Deissler}},
  \bibinfo{author}{\bibfnamefont{C.}~\bibnamefont{D'Errico}},
  \bibinfo{author}{\bibfnamefont{M.}~\bibnamefont{Fattori}},
  \bibinfo{author}{\bibfnamefont{M.}~\bibnamefont{Jona-Lasinio}},
  \bibinfo{author}{\bibfnamefont{S.}~\bibnamefont{Mueller}},
  \bibinfo{author}{\bibfnamefont{G.}~\bibnamefont{Roati}},
  \bibinfo{author}{\bibfnamefont{M.}~\bibnamefont{Inguscio}}, \bibnamefont{and}
  \bibinfo{author}{\bibfnamefont{G.}~\bibnamefont{Modugno}},
  \bibinfo{journal}{Nature Physics} \textbf{\bibinfo{volume}{5}},
  \bibinfo{pages}{586} (\bibinfo{year}{2009}).

\bibitem[{\citenamefont{Pollack et~al.}(2009)\citenamefont{Pollack, Dries, and
  Hulet}}]{Pollack:2009}
\bibinfo{author}{\bibfnamefont{S.~E.} \bibnamefont{Pollack}},
  \bibinfo{author}{\bibfnamefont{D.}~\bibnamefont{Dries}}, \bibnamefont{and}
  \bibinfo{author}{\bibfnamefont{R.~G.} \bibnamefont{Hulet}},
  \bibinfo{journal}{Science} \textbf{\bibinfo{volume}{326}},
  \bibinfo{pages}{1683} (\bibinfo{year}{2009}).

\bibitem[{\citenamefont{Ferlaino et~al.}(2011)\citenamefont{Ferlaino, Zenesini,
  Berninger, Huang, N\"agerl, and Grimm}}]{Ferlaino:2011}
\bibinfo{author}{\bibfnamefont{F.}~\bibnamefont{Ferlaino}},
  \bibinfo{author}{\bibfnamefont{A.}~\bibnamefont{Zenesini}},
  \bibinfo{author}{\bibfnamefont{M.}~\bibnamefont{Berninger}},
  \bibinfo{author}{\bibfnamefont{B.}~\bibnamefont{Huang}},
  \bibinfo{author}{\bibfnamefont{H.-C.} \bibnamefont{N\"agerl}},
  \bibnamefont{and} \bibinfo{author}{\bibfnamefont{R.}~\bibnamefont{Grimm}},
  \bibinfo{journal}{Few-Body Systems} \textbf{\bibinfo{volume}{51}},
  \bibinfo{pages}{113} (\bibinfo{year}{2011}).

\bibitem[{\citenamefont{{Roy} et~al.}(2013)\citenamefont{{Roy}, {Landini},
  {Trenkwalder}, {Semeghini}, {Spagnolli}, {Simoni}, {Fattori}, {Inguscio}, and
  {Modugno}}}]{Roy:2013}
\bibinfo{author}{\bibfnamefont{S.}~\bibnamefont{{Roy}}},
  \bibinfo{author}{\bibfnamefont{M.}~\bibnamefont{{Landini}}},
  \bibinfo{author}{\bibfnamefont{A.}~\bibnamefont{{Trenkwalder}}},
  \bibinfo{author}{\bibfnamefont{G.}~\bibnamefont{{Semeghini}}},
  \bibinfo{author}{\bibfnamefont{G.}~\bibnamefont{{Spagnolli}}},
  \bibinfo{author}{\bibfnamefont{A.}~\bibnamefont{{Simoni}}},
  \bibinfo{author}{\bibfnamefont{M.}~\bibnamefont{{Fattori}}},
  \bibinfo{author}{\bibfnamefont{M.}~\bibnamefont{{Inguscio}}},
  \bibnamefont{and}
  \bibinfo{author}{\bibfnamefont{G.}~\bibnamefont{{Modugno}}},
  \bibinfo{journal}{arXiv:1303.3843}. 

\bibitem[{\citenamefont{Fedichev et~al.}(1996)\citenamefont{Fedichev, Reynolds,
  and Shlyapnikov}}]{Fedichev:1996d}
\bibinfo{author}{\bibfnamefont{P.~O.} \bibnamefont{Fedichev}},
  \bibinfo{author}{\bibfnamefont{M.~W.} \bibnamefont{Reynolds}},
  \bibnamefont{and} \bibinfo{author}{\bibfnamefont{G.~V.}
  \bibnamefont{Shlyapnikov}}, \bibinfo{journal}{Phys. Rev. Lett.}
  \textbf{\bibinfo{volume}{77}}, \bibinfo{pages}{2921} (\bibinfo{year}{1996}).

\bibitem[{\citenamefont{Weber et~al.}(2003)\citenamefont{Weber, Herbig, Mark,
  N\"agerl, and Grimm}}]{Weber:2003}
\bibinfo{author}{\bibfnamefont{T.}~\bibnamefont{Weber}},
  \bibinfo{author}{\bibfnamefont{J.}~\bibnamefont{Herbig}},
  \bibinfo{author}{\bibfnamefont{M.}~\bibnamefont{Mark}},
  \bibinfo{author}{\bibfnamefont{H.-C.} \bibnamefont{N\"agerl}},
  \bibnamefont{and} \bibinfo{author}{\bibfnamefont{R.}~\bibnamefont{Grimm}},
  \bibinfo{journal}{Phys. Rev. Lett.} \textbf{\bibinfo{volume}{91}},
  \bibinfo{pages}{123201} (\bibinfo{year}{2003}).

\bibitem[{\citenamefont{Greene et~al.}(2004)\citenamefont{Greene, Esry, and
  Suno}}]{Greene:2004}
\bibinfo{author}{\bibfnamefont{C.~H.} \bibnamefont{Greene}},
  \bibinfo{author}{\bibfnamefont{B.}~\bibnamefont{Esry}}, \bibnamefont{and}
  \bibinfo{author}{\bibfnamefont{H.}~\bibnamefont{Suno}},
  \bibinfo{journal}{Nuclear Physics A} \textbf{\bibinfo{volume}{737}},
  \bibinfo{pages}{119 } (\bibinfo{year}{2004}).

\bibitem[{\citenamefont{D'Incao et~al.}(2004)\citenamefont{D'Incao, Suno, and
  Esry}}]{DIncao:2004}
\bibinfo{author}{\bibfnamefont{J.~P.} \bibnamefont{D'Incao}},
  \bibinfo{author}{\bibfnamefont{H.}~\bibnamefont{Suno}}, \bibnamefont{and}
  \bibinfo{author}{\bibfnamefont{B.~D.} \bibnamefont{Esry}},
  \bibinfo{journal}{Phys. Rev. Lett.} \textbf{\bibinfo{volume}{93}},
  \bibinfo{pages}{123201} (\bibinfo{year}{2004}).

\bibitem[{\citenamefont{Mehta et~al.}(2009)\citenamefont{Mehta, Rittenhouse,
  D'Incao, von Stecher, and Greene}}]{Mehta:2009}
\bibinfo{author}{\bibfnamefont{N.~P.} \bibnamefont{Mehta}},
  \bibinfo{author}{\bibfnamefont{S.~T.} \bibnamefont{Rittenhouse}},
  \bibinfo{author}{\bibfnamefont{J.~P.} \bibnamefont{D'Incao}},
  \bibinfo{author}{\bibfnamefont{J.}~\bibnamefont{von Stecher}},
  \bibnamefont{and} \bibinfo{author}{\bibfnamefont{C.~H.}
  \bibnamefont{Greene}}, \bibinfo{journal}{Phys. Rev. Lett.}
  \textbf{\bibinfo{volume}{103}}, \bibinfo{pages}{153201}
  (\bibinfo{year}{2009}).

\bibitem[{\citenamefont{Campbell et~al.}(2010)\citenamefont{Campbell, Smith,
  Tammuz, Beattie, Moulder, and Hadzibabic}}]{Campbell:2010}
\bibinfo{author}{\bibfnamefont{R.~L.~D.} \bibnamefont{Campbell}},
  \bibinfo{author}{\bibfnamefont{R.~P.} \bibnamefont{Smith}},
  \bibinfo{author}{\bibfnamefont{N.}~\bibnamefont{Tammuz}},
  \bibinfo{author}{\bibfnamefont{S.}~\bibnamefont{Beattie}},
  \bibinfo{author}{\bibfnamefont{S.}~\bibnamefont{Moulder}}, \bibnamefont{and}
  \bibinfo{author}{\bibfnamefont{Z.}~\bibnamefont{Hadzibabic}},
  \bibinfo{journal}{Phys. Rev. A} \textbf{\bibinfo{volume}{82}},
  \bibinfo{pages}{063611} (\bibinfo{year}{2010}).
  
\bibitem{molecules}
The rapid switch-off ensures that no atoms are converted into molecules when starting on the $a<0$ side of the resonance~\cite{Koehler:2006}.

\bibitem[{\citenamefont{K{\"o}hler et~al.}(2006)\citenamefont{K{\"o}hler,
  G\'oral, and Julienne}}]{Koehler:2006}
\bibinfo{author}{\bibfnamefont{T.}~\bibnamefont{K{\"o}hler}},
  \bibinfo{author}{\bibfnamefont{K.}~\bibnamefont{G\'oral}}, \bibnamefont{and}
  \bibinfo{author}{\bibfnamefont{P.~S.} \bibnamefont{Julienne}},
  \bibinfo{journal}{Rev. Mod. Phys.} \textbf{\bibinfo{volume}{78}},
  \bibinfo{pages}{1311} (\bibinfo{year}{2006}).

\bibitem[{\citenamefont{Smith et~al.}(2011{\natexlab{a}})\citenamefont{Smith,
  Campbell, Tammuz, and Hadzibabic}}]{Smith:2011}
\bibinfo{author}{\bibfnamefont{R.~P.} \bibnamefont{Smith}},
  \bibinfo{author}{\bibfnamefont{R.~L.~D.} \bibnamefont{Campbell}},
  \bibinfo{author}{\bibfnamefont{N.}~\bibnamefont{Tammuz}}, \bibnamefont{and}
  \bibinfo{author}{\bibfnamefont{Z.}~\bibnamefont{Hadzibabic}},
  \bibinfo{journal}{Phys. Rev. Lett.} \textbf{\bibinfo{volume}{106}},
  \bibinfo{pages}{250403} (\bibinfo{year}{2011}{\natexlab{a}}).

\bibitem[{\citenamefont{Smith et~al.}(2011{\natexlab{b}})\citenamefont{Smith,
  Tammuz, Campbell, Holzmann, and Hadzibabic}}]{Smith:2011b}
\bibinfo{author}{\bibfnamefont{R.~P.} \bibnamefont{Smith}},
  \bibinfo{author}{\bibfnamefont{N.}~\bibnamefont{Tammuz}},
  \bibinfo{author}{\bibfnamefont{R.~L.~D.} \bibnamefont{Campbell}},
  \bibinfo{author}{\bibfnamefont{M.}~\bibnamefont{Holzmann}}, \bibnamefont{and}
  \bibinfo{author}{\bibfnamefont{Z.}~\bibnamefont{Hadzibabic}},
  \bibinfo{journal}{Phys. Rev. Lett.} \textbf{\bibinfo{volume}{107}},
  \bibinfo{pages}{190403} (\bibinfo{year}{2011}{\natexlab{b}}).
  
\bibitem{no_wiggles}
For the range of $a$ values relevant here, Efimov physics is not expected to significantly modify this simple scaling~\cite{Zaccanti:2009}.

\bibitem[{\citenamefont{Braaten et~al.}(2003)\citenamefont{Braaten, Hammer, and
  Kusunoki}}]{Braaten:2003}
\bibinfo{author}{\bibfnamefont{E.}~\bibnamefont{Braaten}},
  \bibinfo{author}{\bibfnamefont{H.-W.} \bibnamefont{Hammer}},
  \bibnamefont{and} \bibinfo{author}{\bibfnamefont{M.}~\bibnamefont{Kusunoki}},
  \bibinfo{journal}{Phys. Rev. A} \textbf{\bibinfo{volume}{67}},
  \bibinfo{pages}{022505} (\bibinfo{year}{2003}).



\end{thebibliography}

%

\end{document}